\documentclass[10pt,conference]{IEEEtran}

\IEEEoverridecommandlockouts

\usepackage{amsmath,amsfonts}
\usepackage[ruled,vlined]{algorithm2e}

\usepackage{graphicx}
\usepackage{array}
\usepackage{amssymb}
\usepackage{amsthm}
\usepackage{graphicx}
\usepackage{color}
\usepackage[style=ieee, backend=biber]{biblatex}

\usepackage{tabularx}

\newtheorem{thm}{Lemma}[section]

\begin{document}
\title{Energy-Efficient Multi-Orchestrator Mobile Edge Learning}
\author{Mhd Saria Allahham, Sameh Sorour, \textit{Senior Member, IEEE},  Amr Mohamed, \textit{Senior Member, IEEE},\\ Aiman Erbad, \textit{Senior Member, IEEE} and Mohsen Guizani, \textit{Fellow, IEEE} 
\thanks{Mhd Saria Allahham and Sameh Sorour are with the School of Computing, Queen's University, ON, Canada (e-mail:{20msa7, sameh.sorour}@queensu.ca)}
\thanks{Amr Mohamed and Mohsem Guizani are with College of Engineering, Qatar University, Qatar (e-mail:amrm@qu.edu.qa, mguizani@ieee.org) }
\thanks{Aiman Erbad is with Division of Information and Computing Technology, College
of Science and Engineering, Hamad Bin Khlifa University, Qatar (e-mail: aerbad@hbku.edu.qa)}

}

\maketitle
\begin{abstract}
Mobile Edge Learning (MEL) is a collaborative learning paradigm that features distributed training of Machine Learning (ML) models over edge devices (e.g., IoT devices). 
In MEL, possible coexistence of multiple learning tasks with different datasets may arise. The heterogeneity in edge devices' capabilities will require the joint optimization of  the \textit{learners}-\textit{orchestrator} association and task allocation. To this end, we aim to develop an energy-efficient framework for \textit{learners}-\textit{orchestrator} association and learning task allocation, in which each \textit{orchestrator} gets associated with a group of \textit{learners} with the same learning task based on their communication channel qualities and computational resources, and allocate the tasks accordingly. Therein, a multi-objective optimization problem is formulated to minimize the total energy consumption and maximize the learning tasks' accuracy. However, solving such optimization problem requires centralization and the presence of the whole environment information at a single entity, which becomes impractical in large-scale systems. To reduce the solution complexity and to enable solution decentralization, we propose lightweight heuristic algorithms that can achieve near-optimal performance and facilitate the trade-offs between energy consumption, accuracy, and solution complexity. Simulation results show that the proposed approaches reduce the energy consumption significantly while executing multiple learning tasks compared to recent state-of-the-art methods.


\end{abstract}
\begin{IEEEkeywords}
edge learning, distributed learning, edge networks
\end{IEEEkeywords}

\section{Introduction}

The growing availability of various computing resources and abundant data have pushed the learning algorithms for this data towards the network edge rather than the cloud. Indeed, centralized cloud-based learning paradigms suffer from the huge latency overhead, and such paradigms are impractical for many real-time edge-based applications (e.g., health monitoring and surveillance) \cite{gunduz2020communicate}. This has led to the emergence of the Mobile Edge Learning (MEL) framework \cite{mohammad2019adaptive_propose, zhu2020toward}. MEL is a framework that combines two originally decoupled areas: Mobile Edge Computing (MEC) and Machine Learning (ML), where it distributes and executes learning tasks (i.e., ML models' training) on wireless edge nodes such as IoT devices, while taking into consideration the heterogeneity in these devices' communication and computation capabilities. There are two main components in the MEL framework: 1) \textit{orchestrators} which are responsible for distributing the learning tasks along with aggregating and synchronizing the updates of the tasks, and 2) \textit{learners}, each of which being responsible for training a local ML model using its possessed or received data. MEL is a generalized framework of both Parallelized Learning (PL) and Federated Learning (FL). The former is orchestrator-oriented, where the orchestrator has the whole dataset needed for learning, but it lacks the needed computational resources to do so. Hence, it distributes the ML model with the data across trusted learners to utilize their available resources. On the other hand, the latter is learner-oriented, where each learner has its private data, but the goal is to learn a global model across all the learners in a distributed manner governed by an orchestrator, without sharing the private data of each learner. However, the heterogeneity of the learners' computing capacities causes the so-called "straggler's dilemma" issue, where the distributed learning process is throttled by the learner with the lowest capabilities \cite{cai2020d2d}. Hence, the orchestrator in both cases needs to allocate the tasks according to the learners' capabilities. Task allocation in PL refers to the number of training data samples that will be sent along with the ML model to each learner, whereas in FL, it refers to the number of data samples from the private dataset that will be considered in the local training of each learner.

Several works have addressed the straggler's dilemma in heterogeneous MEL in the literature \cite{mohammad2019adaptive_propose, mohammad2020task}. For instance, in \cite{mohammad2019adaptive_propose}, the authors first aimed to optimize the task allocation along with maximizing the learning accuracy by maximizing the number of local training iterations within one global cycle that has a predefined time limit. The authors then established an energy-aware optimization paradigm in \cite{mohammad2020task}, where the task allocation and the accuracy are optimized according to the time and energy limits of the learners. 
Besides the "straggler's dilemma" issue, the MEL systems face various challenges such as communication and computation overheads, and the limited resources, where numerous works in the literature tackled these challenges from different perspectives. For instance, the authors in \cite{Dinh2020} have proposed a framework for energy-efficient FL over wireless networks, while addressing the resource allocation problem and taking into account the heterogeneous computing and power resources of the learners. Similarly, the authors in \cite{mo2020energy} have aimed for energy-efficient strategies for resource allocation at the edge by optimizing the transmission power, data rates, and the processing frequency of the learners. Moreover, communication-efficient approaches have been proposed in \cite{mills2019communication,Sattler2020}, the communication needed for learning was reduced by applying various compression techniques for the learning task updates. 
To optimize the learning experience of the learners, a generic approach for MEL systems in \cite{Wang2019} has been proposed to optimize the number of local iterations and global cycles such that the learning goal is achieved with limited resources for each learner. This work has been extended in \cite{mohammad2021optimal} to include the optimization of the task size of each learner while considering their heterogeneity. 

While prior works have assumed only one centralized orchestrator either at the cloud server or the edge, and driven by the disadvantage of the limited number of clients each server can access, the authors in \cite{liu2020client} have proposed and developed Hierarchical FL (HFL). HFL is a client-edge-cloud hierarchy, where an edge server act as an orchestrator for the clients connecting to that edge (i.e., learners), and the cloud acts as a global orchestrator for the edge servers. Furthermore, the problem of orchestrator-learner associations has been addressed in \cite{Mhaisen2021, luo2020hfel}, where the edge server (a.k.a orchestrator) gets assigned a set of edge learners based on their connectivity and the data distributions of the learners. 

Nevertheless, most of the previous works and frameworks have either assumed a single learning task managed by either a single orchestrator in the system, or assumed single learning task with multiple orchestrators under a unified governing cloud. 
In this work, we consider a multi-task MEL system to be administered simultaneously by multiple independent orchestrators. 
Examples of such environments are 1) multiple parallel FL jobs on different datasets stored at different groups of learners, or 2) multiple resource-constrained IoT devices parallelizing their learning tasks simultaneously on a set of nearby learners. Moreover, handling simultaneous learning tasks in the same edge environment with the heterogeneity of both learners' resources and channels quality represent a new challenge in multi-task MEL setting, which was not addressed in previous works. 
To this end, we propose energy-efficient learner-orchestrator association and task allocation techniques in this multi-task MEL system, while accounting for the learners' heterogeneity in terms of computation and communication capabilities.
The contributions of this work can be summarized as follows: 
\begin{enumerate}
    \item First, we formulate a multi-objective optimization problem (MOP) for energy-efficient learner-orchestrator association and task allocation, that aims to minimize the total energy consumption in the system and maximize the learning accuracy at the orchestrators.
    \item Being non-convex and NP-hard to solve, we propose an approach that employs an approximate solution to the relaxed and convexified formulated MOP.
    \item To reduce the complexity of the optimization approach, we propose a set of lightweight heuristic algorithms that promote decentralization in the solution and utilize the solution of a simplified version of that formulated MOP. The performance of both proposed approaches is evaluated, analyzed and compared to a recent state-of-the-art technique in MEL.
\end{enumerate}

The rest of the paper is organized as follows: We introduce the system model in Section \ref{sec:sys_model}. Section \ref{sec:form} describes the formulated multi-objective optimization problem, while Section \ref{sec:sol} present the proposed solution approaches. In Section \ref{sec:cmplx} we study the complexity of the proposed approaches. We show the simulation results in Section \ref{sec:sim} before we conclude in Section \ref{sec:conc}.

\begin{figure*}
    \centerline{\includegraphics[scale=0.48]{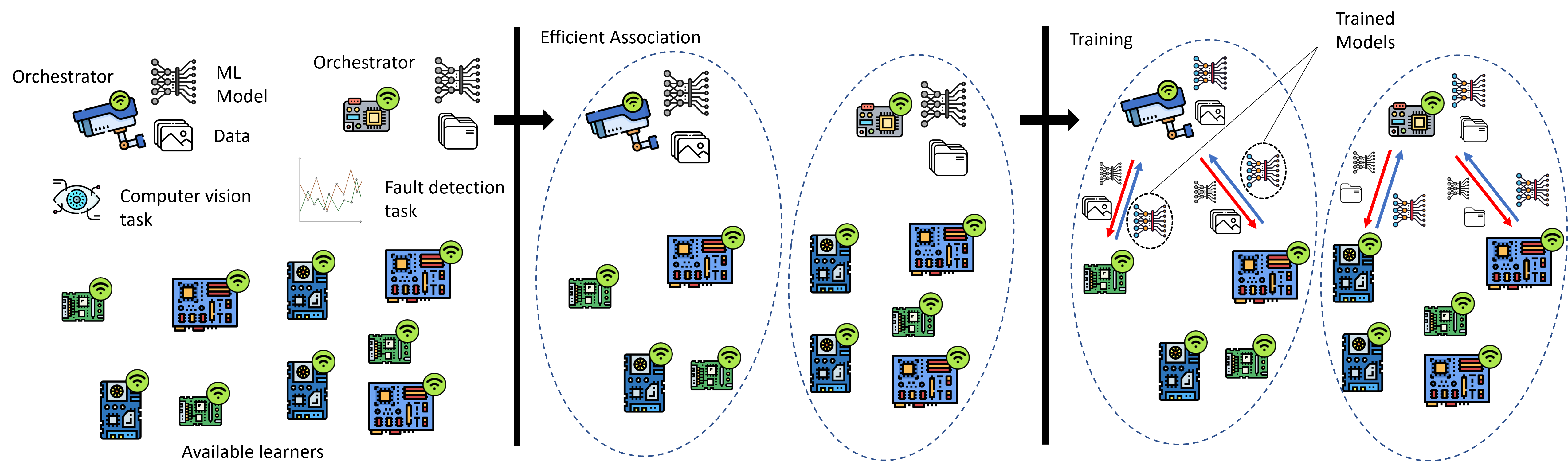}}
    \caption{The considered system model}
    \label{fig:sys_model}
\end{figure*}
\section{System Model}\label{sec:sys_model}
This work considers a multi-task multi-orchestrator MEL system as shown in Fig. (\ref{fig:sys_model}). Each of these orchestrators can be considered as either 1) a governing node for learners that have private data for the same learning problem in the case of FL, or 2) an edge device that lacks the computational resources to execute the training of its learning problem due to its limited capabilities, or the computational resources are exhausted by another task in the case of PL. In this latter case, each of the orchestrators has to distribute its learning task and offload the data needed to the associated learners given their available resources. The learners are considered to be trusted nodes since orchestrators have to share private data with them.
Without loss of generality, we focus in this work on the PL case only in the edge settings, as it can be considered as the general case with the fact that the orchestrators have to offload the data, which is not the case in FL. We will however point out how our formulations and solutions apply to FL whenever needed.

\subsection{Learning Settings}
We denote the set of orchestrators by $\mathcal{O}$, the set of learners by $\mathcal{L}$ and the set of learners that are associated with orchestrator $o$ by  $\mathcal{L}_o$. Each orchestrator $o \in \mathcal{O}$ has a dataset with $N_o$ samples, and each data sample $i$ can be represented by $\{ x_i,y_i \}$, where $x_i$ is the $i^{th}$ feature vector (i.e., input data), and $y_i$ is the $i^{th}$ class or label. After the association with the learners is done, an orchestrator $o$ sends the learning model parameters $\boldsymbol{w}_{l,o}$ and $n_{l,o} \times N_o$ data samples to learner $l$, where $n_{l,o}$ is the proportion of the data to send, and $\sum^{|\mathcal{L}_o|}_{l=1} n_{l,o} = 1$ since the orchestrators need to train on all the available data\footnote{In the FL case, the constraint $\sum_l n_{l,o}=1$ does not necessarily hold since each learner selects a proportion $n_{l,o}$ of its local dataset to perform the learning task. Instead, a constraint has to be added such that the selected samples from a learner represent the distribution of its local dataset.}. Each learner then performs $\tau_o$ local training iterations on its own or received data using Stochastic Gradient Descent (SGD) to minimize its loss function $f_l(\boldsymbol{w}_{l,o})$. Once done, the learners send back to the orchestrator the parameters of their locally trained models, where the latter aggregates these parameters by performing weighted averaging as follows:
\begin{equation}
\boldsymbol{w}_o= \sum_{l \in \mathcal{L}_o}  n_{l,o}\boldsymbol{w}_{l,o}
\end{equation}
Each orchestrator then keeps sending back more data samples and the updated model to the learners, which repeat the same above process for $G_o$ global cycles, until a stopping criteria is satisfied such as the exhaustion of a certain resource (e.g., energy, time) or the attainment of a given accuracy for the aggregated model. One can refer to \cite{Wang2019} for more details about minimizing local loss functions and models aggregations.

\subsection{Mobile Edge Settings}
In this part, we introduce the communication and computation parameters of wireless edge learners. First, we define the number of bits that an orchestrator $o$ sends to learner $l$ as:
\begin{equation}
B^{data}_{l,o} = n_{l,o} N_o F_o\Gamma^d_o
\end{equation}
\begin{equation}
B^{weights}_{o} = S^w_o \Gamma^w_o
\end{equation}
where $F_o$ is the feature vector length, $S^w_o$ the total number of weights in the model, and $\Gamma^d_o$ and $\Gamma^w_o$ represent the bits/feature and bits/weight values, respectively. Therein, we can define the time needed for the orchestrator to send the data and the model weights by:
\begin{equation}
t^S_{l,o} =  \frac{B^{data}_{l,o}+ B_o^{weights}}{W \log_2 (1+ \frac{h_{l,o} P_{l,o}}{\sigma^2})}
\end{equation}
where $W$ is the channel bandwidth, $P_{l,o}$ is the orchestrator's transmission power, $\sigma^2$ is the channel noise variance, and $h_{l,o}$ is the channel gain expressed as $h_{l,o} = d^{-\nu}_{l,o} g^2$ where $d_{l,o}$ is the distance between the orchestrator and the learner, $\nu$ is the path loss exponent, and $g$ is the fading channel coefficient. Similarly, the time needed for a learner to send the updated model parameters is defined as follows:
\begin{equation}
t^U_{l,o} =  \frac{B^{weights}_{o}}{W \log_2 (1+ \frac{h_{l,o} P_{l,o}}{\sigma^2})}
\end{equation}
Last, we define the time needed for a learner $l$ to execute its allocated learning task by:
\begin{equation}
t^C_{l,o} = \frac{\tau_o n_{l,o} N_o C^w_o}{f_l}
\end{equation}
where $\tau_o$ is the number of local iterations, $C^w_o$ is the model computational complexity parameter, and $f_l$ is the local processor frequency. Consequently, the total training time for a learner across $G_o$ global cycles can be expressed as:
\begin{equation}
t_{l,o} = G_o (t^S_{l,o}+t^U_{l,o}+t^C_{l,o})
\end{equation}

We assume in the considered system a fixed bandwidth of $W$, fixed transmission power $P_{l,o} = P_{o,l} = P$ for all the nodes, and $h_{l,o} = h_{o,l}$ (i.e., channel reciprocity). Moreover, we assume that the orchestrator employs a controlled medium access control (MAC) protocol for the learners (e.g., the Reservation protocol) such that the learners when returning back the trained models at the end of the training process are not restricted by a time or channel bandwidth constraints.

Afterward, we define the energy consumption for both learners and orchestrators. Generally, the energy consumed for communications is the product of the transmission power with the transmission time and defined as $E = P\times t$. Therein, the orchestrator $o$ energy consumption to send each cycles’ data samples and the model weights can be given by:
\begin{equation}
E^S_{l,o} = P_{l,o} t^S_{l,o} = \frac{P_{l,o} (B^{data}_{l,o}+ B_o^{weights})}{W \log_2 (1+ \frac{h_{l,o} P_{l,o}}{\sigma^2})}
\end{equation}
Similarly, the learner's $l$ energy consumption to send the updated model is given by:
\begin{equation}
E^U_{l,o} = P_{l,o} t^U_{l,o} = \frac{P_{l,o} B^{weights}_{o}}{W \log_2 (1+ \frac{h_{l,o} P_{l,o}}{\sigma^2})}
\end{equation}
As for the computation energy consumption at each learner $l$, we adopt the model in \cite{mao2017survey_edge_energy}, which can be defined in our context as follows:
\begin{equation}
E^C_{l,o} = \mu \tau_o n_{l,o} N_o C^w_o f_l
\end{equation}
where $\mu$ is the on-board chip capacitance. The total energy consumption for an orchestrator-learner pair during training can thus be expressed as:
\begin{equation}
E_{l,o} = G_o (E^S_{l,o}+E^U_{l,o}+E^C_{l,o})
\end{equation}
To simplify the notation in the reminder of the paper, we define the following time and energy coefficients:
    \begin{description}\small
        \item $A^0_{l,o} = \frac{2  B^{weights}_{o}}{W \log_2 (1+ \frac{h_{l,o} P_{l,o}}{\sigma^2})}$, $\;\;\;\;\zeta^0_{l,o} = P_{l,o} A^0_{l,o}$
        \item $\\$
        \item $A^1_{l,o} = \frac{N_o F_o \Gamma_o^d}{W \log_2 (1+ \frac{h_{l,o} P_{l,o}}{\sigma^2})}$, $\;\;\;\;\zeta^1_{l,o}= P_{l,o} A^1_{l,o}$
        \item $\\$
        \item $A^2_{l,o} =  \frac{N_o C^w_o}{f_l}$, $\;\;\;\;\;\;\;\;\;\;\;\;\;\;\;\;\;\;\;\;\;\;\;\;\;\zeta^2_{l,o}= \mu C^w_o f_l$
    \end{description}
such that the training time and energy consumption can be re-expressed respectively as:
\begin{equation}
t_{l,o} =  G_o (A^2_{l,o} \tau_o n_{l,o} + A^1_{l,o} n_{l,o} + A^0_{l,o}  )
\end{equation}
\begin{equation}
E_{l,o} =  G_o (\zeta^2_{l,o} \tau_o n_{l,o} + \zeta^1_{l,o} n_{l,o} + \zeta^0_{l,o}  )\;
\end{equation}

\section{Problem Formulation}\label{sec:form}
In this section, we first present the distributed learning objective formulation, then present the whole formulation in terms of energy consumption and learning objective in our MEL settings.

\subsection{Learning Objective Formulation}
For a given learners-orchestrator association, aligned with the previous literature \cite{mohammad2021optimal}, we first consider the model presented in \cite{Wang2019} to define the learning objective in the resource constrained system:
\begin{equation}
\begin{matrix}
& \underset{\tau_o, G_o}{\min.} \frac{1}{|\mathcal{O}|}\sum_{o \in \mathcal{O}} F_o(\boldsymbol{w}_o) \\ 
s.t. \\
& G_o (A^2_{l,o} \tau_o n_{l,o} + A^1_{l,o} n_{l,o} + A^0_{l,o}  ) \leq T_{max}, \;\; \forall\; l,o
\end{matrix}
\end{equation}
where $F_o(.)$ is the global loss function for the learning task of an orchestrator $o$, and $T_{max}$ is the maximum allowed training time for the whole learning process. It is generally impossible to find an exact analytical solution for the problem presented in (14) that relates the optimization variables $\tau_o, G_o$ with $F_o(\boldsymbol{w}_o)$. 
Thus, the objective is typically re-formulated as a function of the distributed learning convergence bounds over the edge network. The convergence bounds in the formulation represents how much the trained model in the distributed learning is deviating from the optimal model. Interestingly, it has been shown that the convergence bound mainly depends on the optimization variables, namely, $\tau_o$ and $G_o$. 
The convergence bound for one learning task of one orchestrator has been derived in \cite{Wang2019}, and has been used to reformulate the learning problem objective.
For the sake of brevity, we will only show the important findings that will be included in our main formulation.  
Similar to \cite{Wang2019}, we assume the following about the loss function at each learner $l$:
\begin{enumerate}
    \item $F_l(\boldsymbol{w})$ is convex.
    \item $F_l(\boldsymbol{w})$ is $\beta$-smooth, i.e., $||\nabla F_l(\boldsymbol{w}) - \nabla F_l(\boldsymbol{w}') || \leq  \beta_l || \boldsymbol{w} - \boldsymbol{w}' ||$ for any $\boldsymbol{w}'$.
    \item The divergence between the gradients of the local loss and the aggregated loss function has a maximum of $\delta_l$ such that $||\nabla F_l(\boldsymbol{w}) - \nabla F(\boldsymbol{w}) || \leq \delta_l $
\end{enumerate}
 An auxiliary global model with weights $\boldsymbol{v}_o$ can be then defined, where the auxiliary model represents the centralized training model, which is considered as the optimal training model in the distributed training case. At each global cycle, the model will be updated by SGD as follows:
\begin{equation}
\boldsymbol{v}_o[g_o] = \boldsymbol{v}_o[g_o-1] - \eta_o \nabla F_o(\boldsymbol{v}_o[g_o-1]) 
\end{equation}
where $\eta_o$ is the learning rate and $g_o$ is the index of the global cycles. The difference between the distributed learning weights and the centralized learning at each global update was shown to be upper bounded such that $||\boldsymbol{w}_o[g_o] - \boldsymbol{v}_o [g_o]|| \leq H_o(\tau_o)$, where:
\begin{equation}
H_o(\tau_o) = \frac{\delta_o}{\beta_o} \big[(\eta_o\beta_o+1)^{\tau_o} -\eta_o \delta_o \tau_o\big] 
\end{equation}
where $\delta_o$ can be estimated at each global cycle by $\delta_o = \sum_{l \in \mathcal{L}_o}  n_{l,o} \delta_l$, and $\beta_o$ can be estimated by $\beta_o = \sum_{l \in \mathcal{L}_o}  n_{l,o} \beta_l$, where $\beta_l$ is given as:
\begin{equation}
\beta_l = \frac{||\nabla F_l(\boldsymbol{w}_o) - \nabla F_l(\boldsymbol{w}_l) ||}{|| \boldsymbol{w}_o - \boldsymbol{w}_l ||}
\end{equation}
As a result, the aim of the reformulated objective is to minimize the difference between optimal loss function $F_o(\boldsymbol{w}^*_o)$ (i.e., the centralized loss function $F_o(\boldsymbol{v}_o)$) and the distributed global loss function after $G_o$ global cycles. This difference was shown to be upper bounded as follows \cite{Wang2019}:
\begin{equation}
F_o(\boldsymbol{w}_o) - F_o(\boldsymbol{w}^*_o) \leq \frac{1}{G_o \tau_o \bigg [ \eta_o (1-\frac{\beta_o \eta_o}{2}) - \phi\frac{ H_o(\tau_o)}{\tau_o} \bigg] }
\end{equation}
where $\phi$ is a control parameter. Moreover, the following conditions on the learning rate have to be satisfied to guarantee the convergence:
\begin{enumerate}
    \item $\eta_o \beta_o \leq 1$
    \item $\eta_o (1-\frac{\beta_o \eta_o}{2}) - \phi\frac{ H_o(\tau_o)}{\tau_o} >0$
\end{enumerate}
Even though the aforementioned convergence bound was shown to be convex when $\tau_o \in [1, \tau_{max}]$, where $\tau_{max}$ is the maximum allowed number of local iterations, minimizing over this bound will not have an exact analytical solution. Thus, to enable some solution analysis, we propose to approximate this bound with a simpler convex expression instead. Since the parameters $G_o$ and $\tau_o$ are the only active variables in the expression, and the other variables $\phi, \eta_o, \delta_o$ and $\beta_o$ are either fixed or empirically estimated during training, we can thus define the approximation function as: 
\begin{equation}\small
U_o = \frac{c1}{G_o \tau^{c2}_o}
\end{equation}
where $c1$ and $c2$ are the approximation parameters. Such approximation can be done with log transformation and Linear Regression \cite{changyong2014log_regression}. Consequently, the learning objective is no longer a function of the number of local and global cycles only, but also preserves the convexity of the upper bound with the fact that $\tau_o, G_o \geq 1$. Furthermore, it can be noticed that, as we increase these two parameters, the convergence bound and its approximation in (19) is minimized, which leads to a better learning experience. Finally, although this approximation assumes the loss function is convex, our simulation results show that it also works well in practice for non-convex models (i.e., neural networks).

\subsection{Multi-Objective Optimization Formulation}
Recall that, in this work, we aim to associate each orchestrator $o$ with a set of learners, and to optimize for each orchestrator with its associated learners set  $\mathcal{L}_o$ the task allocation (i.e., $n_{l,o}$, the proportion of the data to be sent to each learner) along with the number of local iterations $\tau_o$, and the number of global cycles $G_o$, such that both the energy consumption in the system and the average global loss functions at all orchestrators are minimized. To formulate this joint problem, we define the MOP as follows:

\begin{subequations}
\begin{align}
& \mathbf{P1}: \;\;\; \underset{\lambda_{l,o}, G_o, \tau_o, n_{l,o}}{\min.}  \alpha \sum_{l,o} \lambda_{l,o} E_{l,o} + (1-\alpha) \sum_{o} U_o\\
&s.t. \sum_o \lambda_{l,o} t_{l,o} \leq T_{max}, \; \forall  \;l \in \mathcal{L}\\
& \quad\;\; \sum_o \lambda_{l,o} = 1, \; \forall  \;l \in \mathcal{L}\\
& \quad\;\; \sum_{l \in \mathcal{L}_o} n_{l,o} = 1, \; \forall \;o \in \mathcal{O}\\
& \quad\;\; 1 \leq \tau_o \leq \tau_{max}, \; \forall \;o \in \mathcal{O}\\
& \quad\;\; \lambda_{l,o} \in \{0,1\}, \; n_{l,o} \in [0,1], \; \forall \;l \in \mathcal{L},\;\;o \in \mathcal{O}\\
& \quad\;\; G_o, \tau_o \in \mathbb{Z^{+}}
\end{align}
\end{subequations}
where $\lambda_{l,o}$ is the association variable between an orchestrator $o$ and a learner $l$, and $\alpha$ is a weighting coefficient that determines the importance of each objective with respect to the other. Constraint (20b) ensures that each learner does not exceed the global time limit for the whole training process, while constraint (20c) guarantees a learner can not associate with more than one orchestrator. Constraint (20d) conveys that the orchestrator offloads the whole dataset to its associated learners, and constraint (20e) guarantees that $\tau_o$ stays in the range where the original convergence bound is convex. Note that the energy and loss objectives in (20a) are averaged over all the learners and orchestrators and normalized between 0-1 by dividing by their maximum values $E_{max}$ and $U_{max}$, respectively, in order to enable a fair trade-off between the objectives.

It is readily obvious that the MOP $\mathbf{P1}$ is a Mixed Non-Linear Integer Program (MNLIP), due to the multiplication and division of the variables in the objective function and the constraints. MNLIP's are known to be non-convex and NP-hard to solve \cite{boyd2004convex}. Usually in such cases, a common approach is to relax the integer variables and consider it as a Geometric Program (GP), but this is not possible in our problem due to the existence of constraints (20c) and (20d). In fact, with the relaxation of the intger variables, the program presented in (20) is an uncommon special case of GP's and known as Signomial Programs (SP) \cite{boyd2004convex}. Generally, SP's are not convex, but several works have proposed algorithms with different approaches to obtain a global solution for them through convexifications and successive solving of sub-optimization problems \cite{floudas2013deterministic, XU2014500_geo_geo, SHEN200499_geo_linear}. In this work, we adopt the presented approach in \cite{SHEN200499_geo_linear} and present some analysis from \cite{floudas2013deterministic}, where we convexify the problem and approximate the non-convex constraints by linear functions, and successively solve relaxed sub-optimization problems until we obtain a global solution. However, such approach is centralized and requires all the information about the environment, the orchestrators and the learners to be available at a single entity, which makes it impractical for large-scale systems. Hence, we further present partially decentralized and fully decentralized light-weight heuristic algorithms that reduce the complexity of the solution. 

\section{Solution Approaches}\label{sec:sol}
In this section, we first present the centralized approach, where we solve the convexified and relaxed version of the presented MOP $\mathbf{P1}$ in (20). Afterward, we present the partially decentralized and fully decentralized approaches to simplify and solve the complex MOP.

\subsection{Centralized Solution using Convex Relaxation}
In order to convexify the problem, we first relax all other integer variables, namely, $\lambda_{l,o}, \tau_o$ and $G_o$, which will be later floored after finding the solution of the simplified convex problem. Since integer variable relaxation expands the feasible region, some undesirable values for the association variables might appear in the solution (e.g., $\lambda_{l,o} = \frac{1}{|\mathcal{O}|}$). Hence, we add the following constraint:
\begin{equation}
\sum^{|\mathcal{O}| - 1}_{i = 1} \sum^{|\mathcal{O}|}_{j = i+1} \lambda_{l,i} \lambda_{l,j} \leq \epsilon,\;\; \; \forall l \in \mathcal{L}
\end{equation}
where $\epsilon$ is a very small number, and $\epsilon > 0$. Along with constraint (20c), this constraint ensures that each learner can only associate with one orchestrator, where its association variable has higher value, and enforces the values of other orchestrators' association variables' to be close to 0.

Afterward, we perform an exponential variable transformation for all the variables such that:
\begin{equation}
x = \exp(\bar{x})
\end{equation}
where $x$ can be either $\lambda_{l,o}, n_{l,o}, \tau_o$ or $G_o$. 
Such transformation will make the multi-objective function in (20a) as sum of exponential terms, which is known to be convex. Consequently, the MOP in $\mathbf{P1}$ can be reformulated as: 
\begin{subequations} \small
\begin{align}
&\begin{split}
\mathbf{P2}:\;\; {\min.}\; \alpha \sum_{l,o}\sum^2_{k=0} \zeta_{l,o}^k \exp(X_k)  &~~~~~~~~~~~~~~~~~~\\
+ (1-\alpha) \sum_{o} c1 \exp(-c2\; \bar{\tau}_o - \bar{G}_o)&
\end{split}\\
&s.t. \sum_o \sum^2_{k=0} A_{l,o}^k \exp(X_k) \leq T_{max}, \; \forall  \;l \in \mathcal{L}\\
& \quad\;\; \sum_o \exp(\bar{\lambda}_{l,o}) -1 \leq 0, \; \forall  \;l \in \mathcal{L}\\
& \quad\;\; 1 - \sum_o \exp(\bar{\lambda}_{l,o}) \leq 0, \; \forall  \;l \in \mathcal{L}\\
& \quad\;\; \sum^{|\mathcal{O}| - 1}_{i = 1} \sum^{|\mathcal{O}|}_{j = i+1}  \exp({\lambda_{l,i}+ \lambda_{l,j}}) \leq \epsilon, \; \forall  \;l \in \mathcal{L}\\
& \quad\;\; \sum_{l \in \mathcal{L}_o} \exp(\bar{n}_{l,o}) -1 \leq 0, \; \forall \;o \in \mathcal{O}\\
& \quad\;\; 1 - \sum_{l \in \mathcal{L}_o} \exp(\bar{n}_{l,o}) \leq 0, \; \forall \;o \in \mathcal{O}\\
& \quad\;\; \bar{\lambda}_{l,o}, \bar{n}_{l,o}, \bar{\tau}_o, \bar{G}_o \in \mathcal{D}
\end{align}
\end{subequations}
where $X_0 = \bar{\lambda}_{l,o}+ \bar{G}_o$, $X_1 = \bar{\lambda}_{l,o}+ \bar{G}_o+ \bar{n}_{l,o}$, $X_2 = \bar{\lambda}_{l,o}+ \bar{G}_o+ \bar{n}_{l,o} + \bar{\tau}_o $, and $\mathcal{D}$ is the new domain of the MOP after the reformulation. This new domain can be defined by applying the transformation on the upper and lower bounds for each variable in constraints (20e)-(20g). It can be noticed that the reformulated MOP $\mathbf{P2}$ is convex in the objective and the constraints, except for constraints (23d) and (23g), where there exist concave terms. Hence, 
by utilizing the following linear function:
\begin{equation}\small
L(x) = \frac{x_{max} e^{x_{min}} - x_{min} e^{x_{max}}} {x_{max} - x_{min}} + \frac{ e^{x_{max}} - e^{x_{min}}}{x_{max} - x_{min}} x
\end{equation}
where $x_{min}, x_{max}$ represent the bounds for the variables $\lambda_{l,o}$ and $n_{l,o}$ in $\mathbf{P1}$ after the integer relaxation. We relax the MOP $\mathbf{P2}$ by  underestimating each of the exponential terms in constraints (23d) and (23g) such that the constraints can be rewritten as affine functions as follows:
\begin{subequations}
\begin{align}
& 1 -  \sum_o L(\bar{\lambda}_{l,o}) \leq 0\\
& 1 -  \sum_{l \in \mathcal{L}_o} L(\bar{n}_{l,o}) \leq 0
\end{align}
\end{subequations}
In fact, the linear underestimator represents a lower bound on the concave terms in (23d) and (23g), and the smaller the difference between the underestimation and the concave terms the closer the solution of the relaxed MOP will be to the solution of the MOP in (23). By making use of the analysis shown in \cite{floudas2013deterministic}, we can assess the quality of this lower bounding by examining the tightness of the underestimation of every concave term with the linear function (24) inside an interval $[x_{min}, x_{max}]$. 
\begin{thm}
Given the separation function $\Delta(x)$ as the difference between the concave term and its underestimator, $\Delta(x)$ is concave in $x$ and its maximum can be given by:
\begin{equation}
\Delta_{max} = e^{x_{min}} \left( 1 - Z + Z \log(Z)    \right) 
\end{equation}
where
\begin{equation}
Z = \frac{e^\vartheta - 1}{\vartheta}, \; \vartheta = x_{max} - x_{min}
\end{equation}
\end{thm}

\textit{Proof.} The proof is detailed in Appendix \ref{appA}. We follow the same presented procedure in \cite{floudas2013deterministic} with a slight modification.

It can be noticed that as the interval $\vartheta$ goes to zero, $Z$ approaches one, and hence the maximum separation goes to zero. 
Furthermore, the rate at which $\Delta_{max}$ goes to zero can be examined using the Taylor series expansion of (26)\footnote{Taylor series expansion of (26) can be either derived by starting with $\frac{e^x - 1}{x} = \sum^\infty_{n=0} \frac{x^n}{(n+1)!}$ and finding the derivatives of the other terms, or by using software such as MATLAB.} as follows:
\begin{equation}
\frac{\Delta_{max}}{e^{x_{min}}} = \frac{\vartheta^2}{8} + \frac{\vartheta^3}{16} + \frac{11\vartheta^4}{576}+ \frac{5\vartheta^5}{1152} + O(\vartheta^6)
\end{equation}
Taking into consideration the first term only, it can be deduced that the rate at which $\Delta_{max}$ goes to zero is:
\begin{equation}
\Delta_{max} \approx O(\vartheta^2), \;\; as \;\;\vartheta \rightarrow 0.
\end{equation}
We can conclude that if an effective relaxation is desired, a sufficiently small difference between the lower and the upper bounds is required. 

After the convexification of $\mathbf{P2}$, a Branch and Bound (BnB) algorithm is employed as in \cite{SHEN200499_geo_linear}. This approach solves a sequence of convex sub-problems of $\mathbf{P2}$ over partitioned subsets of $\mathcal{D}$ in order to obtain a global optimum solution. The BnB approach consists of $k$ stages, where in each stage the set $\mathcal{D}^k$ is partitioned into subsets, each concerned with a node in the BnB-tree, and each node is associated with a sub-problem of $\mathbf{P2}$ in each subset. In each stage, the feasibility of each sub-problem in each node is checked and solved via interior point methods to obtain a lower bound on the optimal value of $\mathbf{P2}$. Subsets that obtain a better lower bound than the previous stage  are then partitioned again, each with a new node. This process is repeated until convergence, or the maximum number of stages is reached. Interested readers can refer to \cite{SHEN200499_geo_linear} for the full algorithm details and proof of convergence. Such BnB approaches for solving SP's are already available in optimization solvers such as GPkit \cite{burnell2020gpkit}. Nevertheless, it is important to note that, while obtained solutions from the aforementioned approach can be optimal for the non-convex $\mathbf{P2}$, but it might not be the case for $\mathbf{P1}$ due to the integer variable relaxation.

\subsection{Partially Decentralized Heuristics}
In this work, we propose partially decentralized heuristics, where orchestrators need to cooperate and communicate by sharing the information (i.e., dataset size, channel qualities with learners...etc.) to realize the association with the learners. Once the association is done, each orchestrator can optimize the task allocation, the number of local training iterations for each associated learner, and the number of the global cycles. 
\subsubsection{\textbf{The Associate-Allocate-Train Decomposition Approach}}
First, we propose the Associate-Allocate-Train (AAT) algorithm, where the MOP $\mathbf{P1}$ is broken down into three simpler sub-problems as follows:
\begin{equation}
\begin{split}
&\textbf{SP1}:~~ \underset{\lambda_{l,o}}{\min.}~~\sum_{l,o} \lambda_{l,o} E_{l,o}~~~~~~~~~~~~~~~\\ &s.t. ~~ \text{(20b), (20c)}, ~ \lambda_{l,o} \in \{0,1\}  
\end{split}
\end{equation}
\begin{equation}
\begin{split}
 &\textbf{SP2}:~~ \underset{n_{l,o}}{\min.}~~\sum_{l\in \mathcal{L}_o} n_{l,o} G_o (  \zeta^2_{l,o} \tau_o + \zeta^1_{l,o} ) \\ &s.t. ~~ \text{(20b), (20d)}, ~ n_{l,o} \in [0,1]
\end{split}
\end{equation}

\begin{equation}
\begin{split}
 &\textbf{SP3}:~~ \underset{\tau_o, G_o}{\min.}~~ \alpha \sum_{l\in \mathcal{L}_o} E_{l,o} + (1-\alpha) \frac{c1}{G_o \tau_o^{c2}}  \\& s.t. ~~ \text{(20b), (20e), (20g)}
\end{split}
\end{equation}
This decomposition is driven by the fact that both task allocation and training are dependent on the association. Hence, in the first sub-problem \textbf{SP1}, the orchestrators assume equal task allocation for all the learners (i.e., $n_{l,o} = \frac{1}{|\mathcal{L}|}$) and a fixed number of local iteration and global cycles so that the energy consumption for each association is known. \textbf{SP1} then optimizes the associations such that the total energy consumption is minimized. One can clearly see that \textbf{SP1} is a binary integer Linear Program  (LP), and efficient methods to obtain global solutions to such programs already exist in the literature  \cite{kochenderfer2019algorithms, fast_integer}. After obtaining the association variables, the task allocation sub-problem \textbf{SP2} can be solved. In fact, \textbf{SP2} is a simple LP that can be solved efficiently by each orchestrator to determine the task size $n_{l,o}$ for each learner. Finally, each orchestrator is left with the training sub-problem \textbf{SP3} to solve, which determines the number of local iterations and global cycles and also controls the energy-accuracy trade-off. Since \textbf{SP3} contains only two integer variables, we can employ exhaustive search to find the optimal values for $\tau_o$ and $G_o$. However, to achieve a faster search, we opt to find an optimal upper bound on both variables in order to reduce the search space. First, within a group of associated learners, we express $l^*$ as the learner index with the maximum training time in that group such that $l^* = \arg ~ \underset{l \in \mathcal{L}_o}{\max}~~ t_{l,o}$.
\begin{thm}
For $c2 = 1$ and $G_o \in [1, \frac{1}{\xi})$, the optimal upper bounds for $G_o$ and $\tau_o$ can be given by:
\begin{equation}
G^{{max}^*}_o = \left \lfloor \frac{1- \sqrt{\frac{\xi a \theta^2}{b\xi - \theta c}} }{\xi } \right \rfloor
\end{equation}
\begin{equation}
\tau^{{max}^*}_o = \min \left ( \left \lfloor \frac{ 1 - \xi G^{{max}^*}_o  }{ \theta G^{{max}^*}_o } \right \rfloor, \tau_{max} \right )
\end{equation}
where 
    \begin{description}
        \item $a = \frac{(1-\alpha)c1}{U_{max}},~~ b = \frac{\alpha  \sum_l \zeta^2_{l^*,o} n_{l^*,o} }{E_{max} |\mathcal{L}_o| }$
        \item $\\$
        \item $c = \frac{\alpha  \sum_l \left ( \zeta^1_{l^*,o} n_{l^*,o} + \zeta^1_{l^*,o} \right )}{E_{max} |\mathcal{L}_o| },~~ \theta= \frac{A^2_{l^*,o} n_{l^*,o}}{T_{max}}$
        \item $\\$
        \item $\xi =  \frac {(A^1_{l^*,o} n_{l^*,o} + A^0_{l^*,o}  )}{T_{max}}$
    \end{description}
\end{thm}

\textit{Proof.} The proof is detailed in Appendix \ref{appB}.

Lemma 2 shows that the maximum number of global cycles and the maximum local iterations are both dependent on the learner with least capabilities, which thus takes the longest training time. Moreover, in practice, the regression parameter $c2$ depends on the learning parameters (i.e., $\beta_o$ and $\eta_o$). We can thus set empirical upper bounds on these values which set the parameter $c2$ to 1.
As a result, the optimal values for the number of local iterations and global cycles can be found by searching over the intervals $[1, \tau^{max}_o]$ and $[1,G^{max}_o]$ for $\tau_o$ and $G_o$, respectively. 

Nevertheless, the two sub-problems \textbf{SP2} and \textbf{SP3} in the above heuristics are coupled together as the number of iterations and global cycles are optimized based on how much data each learner received, and vice versa, such that the training time does not exceed the limit. Moreover, solving each sub-problem separately might result in inefficient task allocation and poor choices for the number of local iterations and global cycles. As such, we propose an iterative procedure for jointly optimizing \textbf{SP2} and \textbf{SP3} by repeatedly alternating the minimization over the task allocation variables on one side, and the number of local iterations and global cycles on the other side. This procedure is as follows: 1) Initialize values for $\tau_o$ and $G_o$, solve the task allocation LP in \textbf{SP2} to obtain the task size for each learner; 2) Determine the optimal values for $G_o$ and $\tau_o$ according to \textbf{SP3} by searching over the intervals that are defined by (33) and (34); 3) Repeatedly alternate the optimization between the two sub-problems until convergence (i.e., until the objective value of \textbf{P1} converges). The full details of the AAT algorithm are provided in Algorithm 1.

 \begin{algorithm}[t]\label{algo:AAT_algo}
\SetAlgoLined
 Initialize: $\alpha, \;T_{max}, \tau_{max}, \; \tau_o, \; G_o$\\
\textbf{At each orchestrator $o$ :}\\
Assume equal task allocation for all the learners $n_{l,o} = \frac{1}{|\mathcal{L}|}$\;
Acquire other orchestrators information\;
Solve the association problem (30)\;
Obtain the set of associated learners $\mathcal{L}_o$\;
\While{no convergence}{
Solve the task allocation problem in (31)\;
Obtain the optimal upper bounds $G^{{max}^*}_o$ and $\tau^{{max}^*}_o$ according to (33) and (34)\;
Perform exhaustive search to solve (32) and find the optimal number of iterations and global cycles\;
}
\textbf{return} $n_{l,o}, \tau_o, G_o$
 \caption{Assign-Allocate-Train (AAT) heuristic}
\end{algorithm}

\subsubsection{\textbf{Factor-Based Association and Allocation}}
We present another heuristic approach that we refer to as the Factor-Based Association and Allocation (FBA). The FBA is based on an association factor (AF) $\Lambda_{l,o}$, which is expressed as:
\begin{equation}
\Lambda_{l,o} = \frac{\bar{f}_{l}}{\bar{d}_{l,o}}
\end{equation}
where $\bar{f}_{l}$ and $\bar{d}_{l,o}$, both $\in [0,1]$, are the normalized processor frequency of learner $l$ and the distance between learner $l$ and orchestrator $o$, respectively. The AF characterizes each learner's computing capability and its distance-based connection quality to each orchestrator. Each orchestrator can obtain this factor from all the available learners and share it with other orchestrators. Based on these AFs, the FBA first performs a centralized turn-based association, where in each turn an orchestrator is selected fairly to get its chance to associate with a learner, and the learner with the maximum AF to that orchestrator gets associated with it. After an orchestrator is aware of its associated learners, each learner will be allocated a task size according to the following:
\begin{equation}
n_{l,o} = N_o \times \frac{\Lambda_{l,o}}{\sum_{l \in \mathcal{L}_o} \Lambda_{l,o}}
\end{equation}
The rationale behind such task allocation is that learners with higher AF will get allocated larger amounts of data samples for training. Indeed, such learners will execute their tasks faster and have better channel quality for data and model transmissions with closer orchestrators. After determining the task size for each learner, orchestrators can do the same exhaustive search to solve (32) over the intervals that can be defined by (33) and (34) in order to obtain the optimal number of local iterations and global cycles. The FBA algorithm is detailed in Algorithm 2.

 \begin{algorithm}[t]\label{algo:FBA_algo}
\SetAlgoLined
Initialize: $\alpha, \;T_{max},\; \tau_{max},\; \mathcal{O},\; \mathcal{L}$\\
Obtain from the learners their AFs $\Lambda_{l,o}$\;
\While{$\mathcal{L} \neq \varnothing  $}{
Initialize a buffer for the orchestrators $\mathcal{O}_{temp} \leftarrow \mathcal{O}$\;
\While{$\mathcal{O}_{temp} \neq \varnothing ~\&~ \mathcal{L} \neq \varnothing$}{
Select an orchestrator $O_i$ randomly from $\mathcal{O}_{temp}$ \;
Associate it with learner $L_j$ where:  $ j =  \arg \underset{l}{\max} \;\; \Lambda_{l,o}$  \;
Remove the associated learner from the set of available learners: $\mathcal{L} \leftarrow \mathcal{L}\;\setminus L_j$ \;
Remove the associated orchestrator from the buffer: $\mathcal{O}_{temp} \leftarrow \mathcal{O}_{temp}\;\setminus O_i$ \; } }
\textbf{At each orchestrator:} \\
Perform the task allocation according to (36)\;
Obtain the optimal upper bounds $G^{{max}^*}_o$ and $\tau^{{max}^*}_o$ according to (33) and (34)\;
Perform exhaustive search to solve (32) and find the optimal number of iterations and global cycles\;
\textbf{return} $n_{l,o}, \tau_o, G_o$
 \caption{Factor-Based Association (FBA) heuristic}
\end{algorithm}

\subsection{Fully Decentralized Heuristic}
Lastly, we propose a fully decentralized approach, namely, the Learner-driven FBA (L-FBA). In L-FBA algorithm, the learners initiate the association by calculating their AFs for each orchestrator, selecting and associating with the orchestrator with the highest AF, and informing their selected orchestrators about their AF value. After an orchestrator receives the list of associated learners along with their AF values, it can determine the task sizes based on each learner AF similarly to the original FBA algorithm according to the task allocation equation (36). Finally, the orchestrator can perform the same exhaustive search with the specified bounds to find the optimal number of local iterations and global cycles. The L-FBA algorithm is summarized in Algorithm 3.
\begin{algorithm}\label{algo:LFBA_algo}
\SetAlgoLined
Initialize: $\alpha, \;T_{max},\; \tau_{max}$\\
\textbf{At each learner:}\\
Look for the available orchestrators\;
Calculate the AF for each orchestrator $\Lambda_{l,o}$ \;
Associate with an orchestrator $O_i$ where:
$ i =  \arg \underset{o}{\max} \;\; \Lambda_{l,o}$  \;
\textbf{At each orchestrator:}\\
Receive the list of associated learners with their AF's $\Lambda_{l,o}$\;
Perform the task allocation according to (36)\;
Obtain the optimal upper bounds $G^{{max}^*}_o$ and $\tau^{{max}^*}_o$ according to (33) and (34)\;
Perform exhaustive search to solve (32) and find the optimal number of iterations and global cycles\;
\textbf{return} $n_{l,o}, \tau_o, G_o$
 \caption{Learner-driven FBA (L-FBA) heuristic}
\end{algorithm}

\section{Complexity Analysis}\label{sec:cmplx}
In this section, we study the complexity of each proposed algorithm, namely, the centralized optimization (COPT), the AAT, the FBA and the L-FBA algorithms. For the ease of reading, some variables will be reused and redefined in this section.

\subsection{The COPT Algorithm:}
First, the COPT employ the BnB algorithm to find a global solution, which worst case complexity is known to be $O(b^k)$, where $b$ is the number of branches per node, and $k$ is the maximum number of iterations for the BnB algorithm, i.e., the depth of the constructed BnB tree. In each node, the COPT solves a convex sub-problem using the interior point method, which has a complexity of $O(\sqrt{n}\log\left(\frac{\mu_0 n}{\varepsilon}\right) )$, where $n$ is the dimension of the domain $\mathcal{D}$ and expressed in our problem as $n = 2|\mathcal{O}| (|\mathcal{L}| + 1)$, $\varepsilon$ is the tolerance variable, and $\mu_0$ is a hyper parameter \cite{interioir_p}. Thus, the complexity of the COPT is given by $O(\sqrt{n}\log\left(\frac{\mu_0n}{\varepsilon}\right)\times b^k)$. 

\subsection{The AAT Algorithm:}
As for the AAT, we first analyze each sub-problem complexity. For \textbf{SP1}, an integer linear program is solved, which has complexity $O( c + \log \left(  c  \right) \rho)$, where $c$ is the number of constraints in the program, and $\rho$ is the bit precision hyperparameter \cite{fast_integer}. In fact, the number of constraints for \textbf{SP1} is $c = 2|\mathcal{L}| $.  The sub-problem \textbf{SP2} is a simple linear program, and its solving complexity is given by $O(\mathcal{C}\,\sqrt{c})$, where $\mathcal{C}$ represents the bit complexity  \footnote{Bit complexity is the number of single operations (of addition, subtraction, and multiplication) required to complete an algorithm.} \cite{simple_LP}. The complexity of \textbf{SP3} is the complexity of the exhaustive search and can be expressed as $O(\tau_{max}  G_{max})$, where $\tau_{max}$ and $G_{max}$ represents the upper bounds of the search interval. As a result, the overall complexity for the AAT algorithm can be given by $O( c + \log \left(  c  \right) \rho + k( \mathcal{C}\,\sqrt{c} + \tau_{max}  G_{max}  ) )$.

\subsection{The FBA and L-FBA Algorithms:}
The FBA algorithm first associates each learner to an orchestrator, then each orchestrator performs the task allocation for each associated learner, and the exhaustive search to obtain the number of local iterations and global cycles. Hence, its complexity can be expresses as $O( 2|\mathcal{L}| + \tau_{max}  G_{max})$. Lastly, the L-FBA approach is totally decentralized, where learners only do basic operations to determine the AF for each orchestrator, and then the orchestrator do the task allocation and the exhaustive search. Therefore, its complexity is given by $O( |\mathcal{L}| + \tau_{max}  G_{max})$.

One can see that the COPT approach is the most complex approach. In fact, as the number of learners and orchestrators increase in the system, its complexity can grow exponentially. This is due to the fact that the number of needed iterations or branches for the BnB algorithm can increase as the dimensions of the domain increase, which is not the case for the other algorithms. On the other extreme, the FBA and L-FBA are the least complex ones, as the orchestrator is only required to do the task allocation and search for the optimal number of iterations and global cycles, and their complexity scale linearly as the number of learners increases.

\section{Simulation Results}\label{sec:sim}
In this section, we first show the energy-accuracy trade-offs for the proposed algorithms, and then compare our approaches with the recent state-of-art MEL approach in \cite{mohammad2021optimal}, which we will refer to as the Energy-Unaware (EU) approach. Afterward, we show the performance of the algorithms in different scenarios where we: 1) Fix the number of orchestrators and vary the number of learners; 2) Vary the number of orchestrators and fix the number of learners in the system. The comparison and the performance evaluation are done considering 3 orchestrators and 50 learners, where the orchestrators have similar multiple learning tasks (i.e., similar datasets and models architecture) for the sake of convenience in the results. We first utilize the MNIST dataset for performance comparison and evaluation. Subsequently, we evaluate the learning performance considering multiple and different tasks and datasets. The simulations were run considering the parameters shown in Table \ref{table:parameters}. The learning parameters values are taken from \cite{Wang2019} as is, and the architectures for the models used can be found in Appendix \ref{appC}.
\begin{table}[t!]
\caption{Simulation parameters}
 \label{table:parameters}
\centering
\begin{tabular}{ll}
\hline
Node Bandwidth $W$  &       5 MHz\\ \hline
Transmission Power $P_{l,o}$ &      200 mW \\ \hline
Distance range for $d_{l,o}$ &           $[5,50]$ m \\ \hline
Processors frequencies $f_l$'s  &    $[0.5, 0.7, 1.2, 1.8]$ GHz \\ \hline
On chip capacitance $\mu$ &     $1\times10^{-19}$\\ \hline
Learning parameters $\eta_o,\phi$ & [0.01,0.0001] \\ \hline
Maximum weights divergence $\delta_o$ & 5 \\ \hline
Maximum gradients divergence $\beta_o$ & 0.5 \\ \hline
Bit precision for $\Gamma^w_o$ and $\Gamma^d_o$ & 32 bits\\ \hline
Dataset size for all datasets & 60,000\\ \hline
\end{tabular}
\end{table}

\begin{figure}
    \centering
    \includegraphics[scale = 0.5]{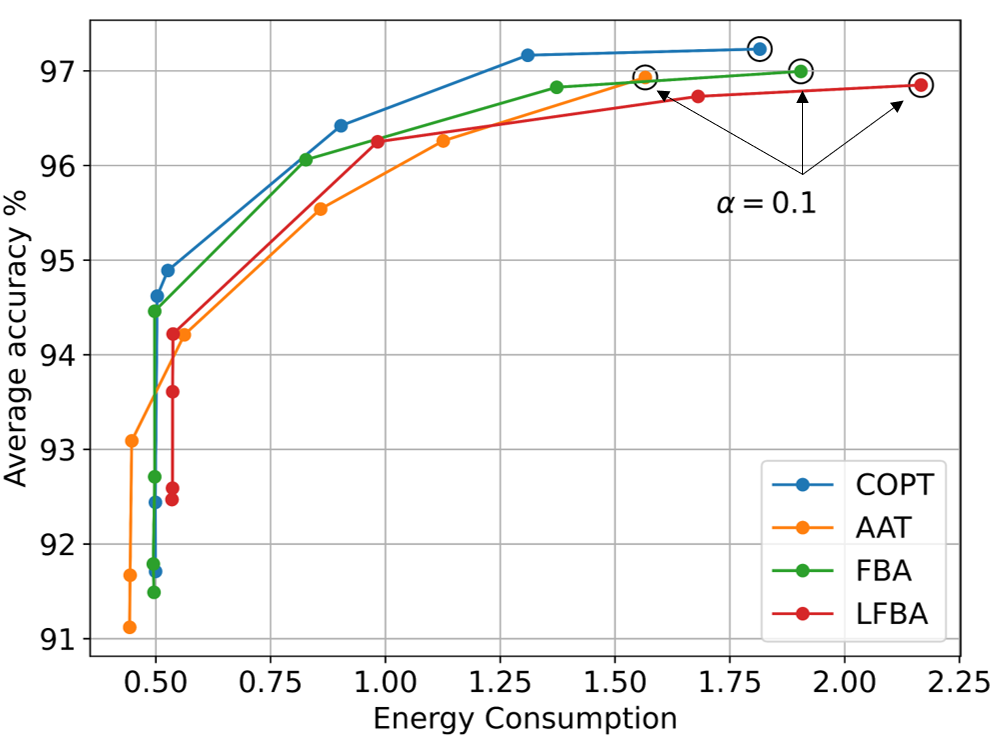}
    \caption{Energy-accuracy trade-offs curves considering $T_{max} = 660s$ and }
    \label{fig:my_pareto}
\end{figure}
\subsection{Energy-Accuracy Tradeoff}
Objectives Pareto trade-off curves are generated using different values for the MOP weight, namely, $\alpha$. 
These curves help us to find Pareto optimal solutions for the weights such that it can balance the performance between the energy consumption and the accuracy. 
Each point on the curves represents a solution for a single value of the weight.
The trade-off curves for the proposed algorithms are shown in Fig. \ref{fig:my_pareto}. We can see that the COPT approach achieves the best trade-off as it achieves the highest accuracy with a low energy consumption. Moreover, the AAT approach is the best energy conservative approach, but it underperforms in terms of accuracy. This is due to the fact that the AAT algorithm optimizes the association and the task allocation energy consumption only first, and then it considers the energy-accuracy trade-off when deciding the number of local iterations and global cycles. As for the FBA, it performs slightly better than the LFBA approach, but they have similar performance where they have better accuracy than the AAT and worse than the COPT, but more energy consumption in general.
Lastly, it can be noticed the Pareto optimal values will lay in the range of $\alpha$ between 0.2 and 0.4, where decreasing the weight will increase the accuracy without significantly increasing the energy consumption. 

\begin{figure}
    \centering
    \includegraphics[scale = 0.55]{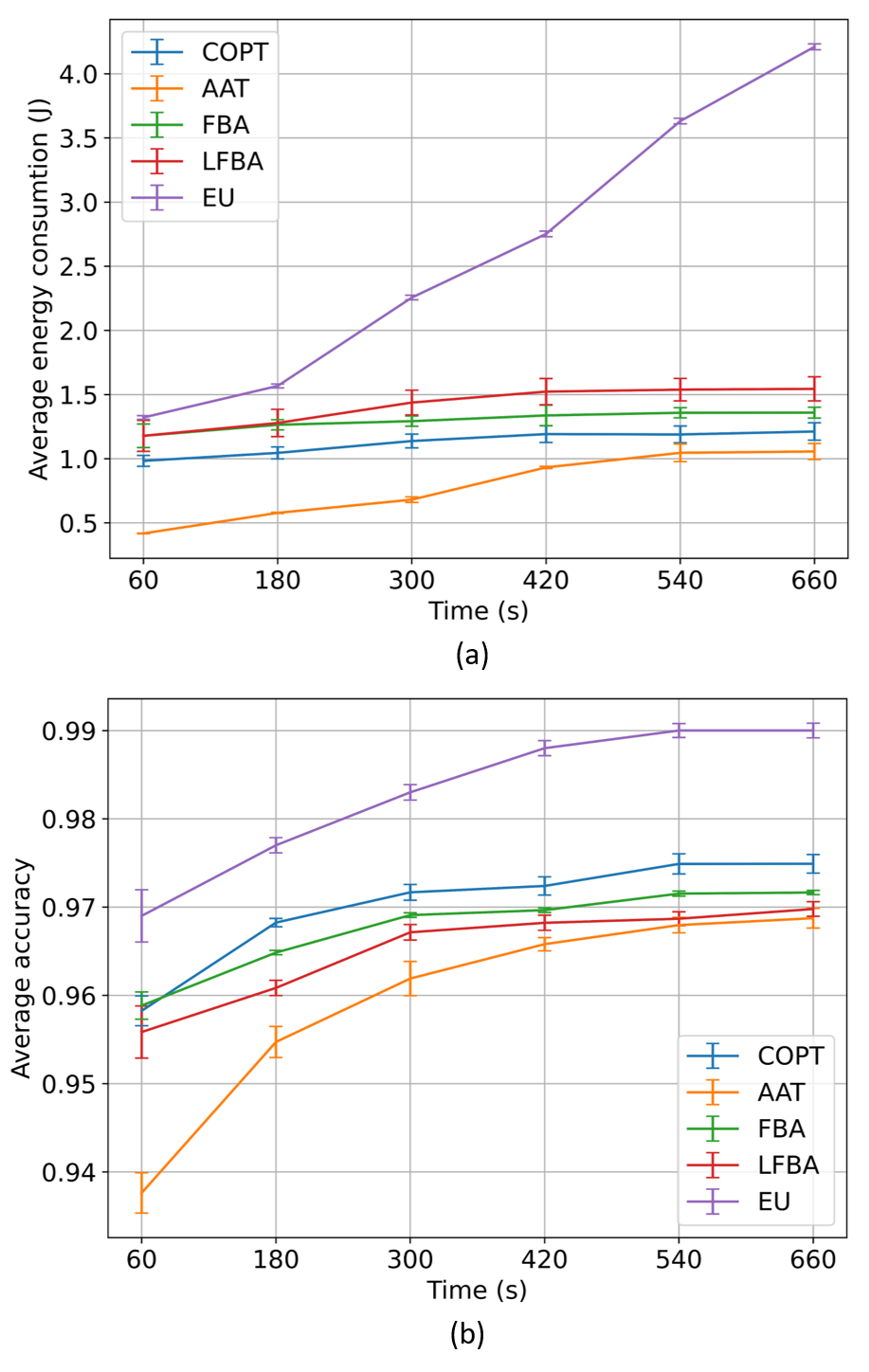}
    \caption{Performance comparison between approaches in terms of (a) energy consumption (b) learning accuracy with 3 orchestrators and 50 learners}
    \label{fig:my_comp}
\end{figure}
\subsection{Performance Comparison with the EU Approach}
We compare our approach with the EU technique that is presented in \cite{mohammad2021optimal} with distance based association. The EU approach optimizes the task allocation and the number of training iterations between heterogeneous learners such that the learning experience is maximized under global time constraints. We have conducted the comparative study using Monte Carlo simulation with 50-100 runs. The comparisons in terms of energy consumption and accuracy are depicted in Fig. \ref{fig:my_comp} (a) and (b), respectively, considering different values of the time constraint $T_{max}$ . 
In Fig. \ref{fig:my_comp} (a), we can see that, as we increase $T_{max}$, all approaches consumes more energy, since increasing $T_{max}$ adds more degree of freedom to do more local iterations and global cycles. However, the energy consumption for all the proposed approaches is significantly lower than the EU approach, and the optimization approach consumes slightly less energy than both the FBA and LFBA approaches, while the AAT approach consumes the least energy in the system.
Fig. \ref{fig:my_comp} (b) shows that the EU approach achieved the highest accuracy, but the proposed COPT approach only falls behind within 2\% accuracy range for the different values of $T_{max}$. On the other hand, the proposed heuristics underperform for smaller $T_{max}$ values, but it only falls behind within 3\% range for larger $T_{max}$ values. 

\begin{figure}
    \centering
    \includegraphics[scale = 0.55]{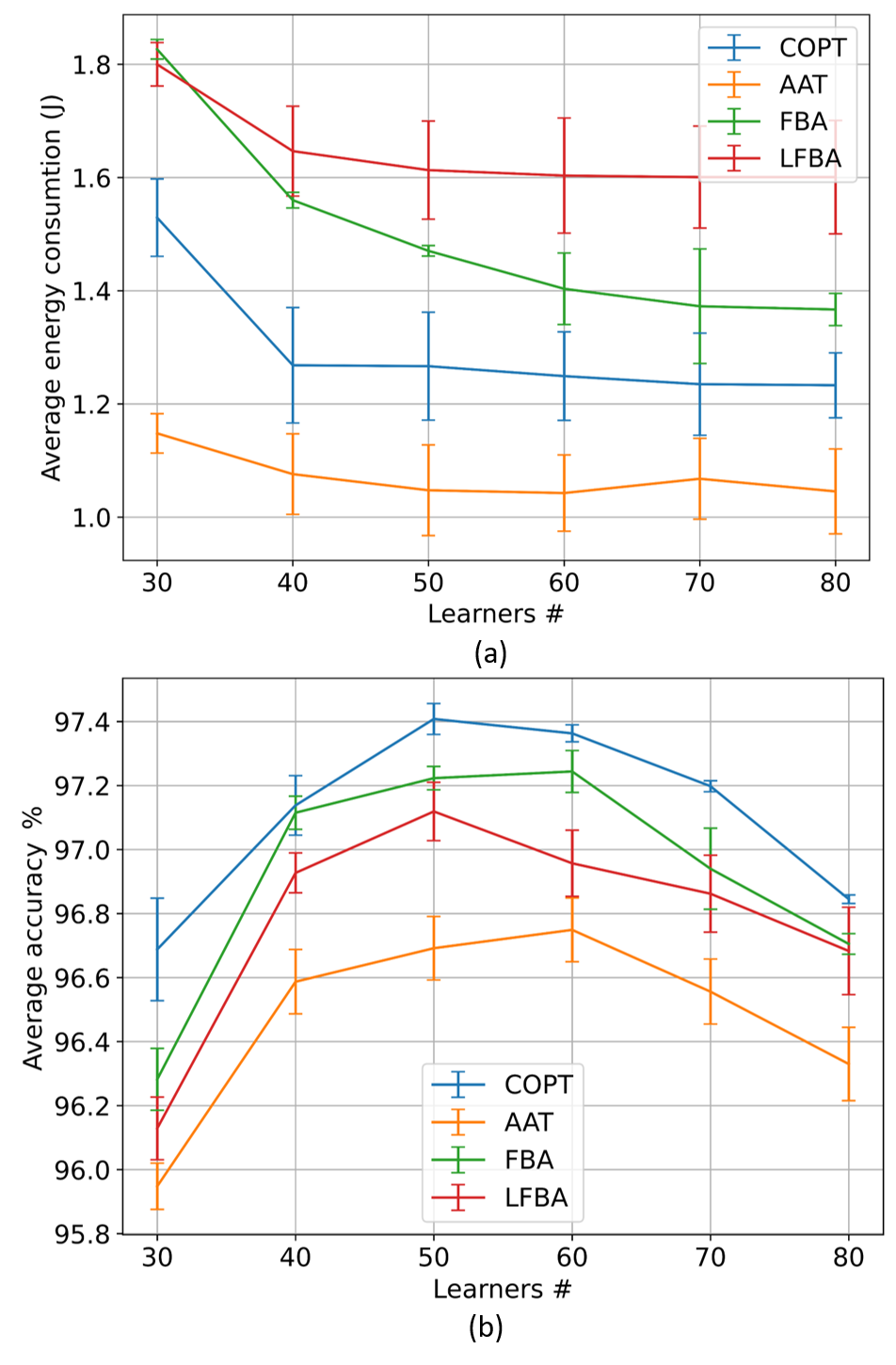}
    \caption{Performance evaluation with different number of learners in terms of (a) energy consumption (b) learning accuracy while considering 3 orchestrators and $T_{max} = 660s$.}
    \label{fig:my_eval_learners}
\end{figure}

\begin{figure}
    \centering
    \includegraphics[scale = 0.55]{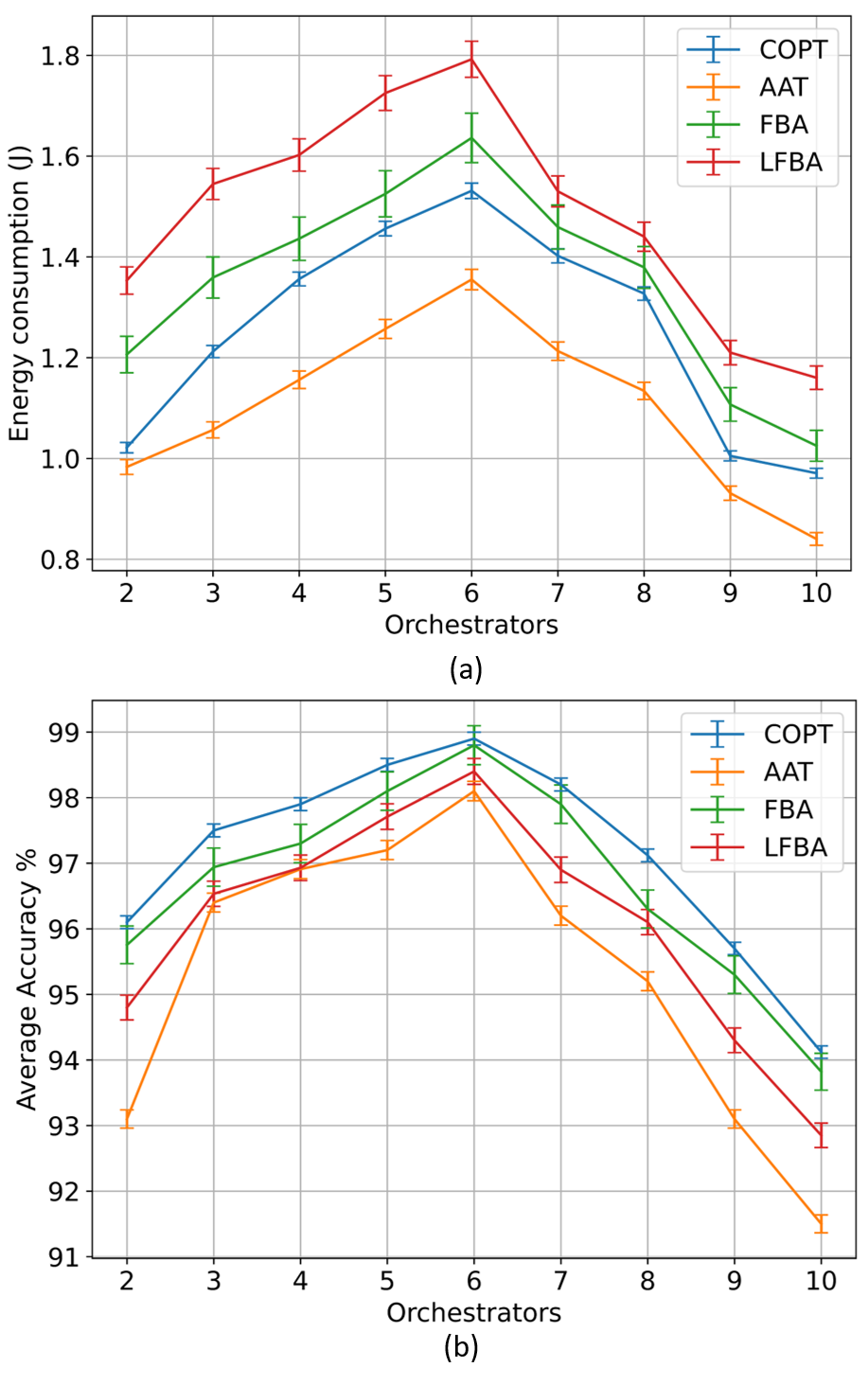}
    \caption{Performance evaluation with different number of orchestrators in terms of (a) energy consumption (b) learning accuracy while considering 50 learners and $T_{max} = 660s$.}
    \label{fig:my_eval_orchs}
\end{figure}

\subsection{Performance Evaluation in Different Scenarios}
We first show the performance of our proposed algorithms in terms of energy consumption and accuracy when varying the number of learners while fixing the number of orchestrators in Fig. \ref{fig:my_eval_learners}. Then in Fig. \ref{fig:my_eval_orchs} we show the performance when varying the number of orchestrators while fixing the number of learners. In Fig. \ref{fig:my_eval_learners} (a), we can see that as the number of learners is increased, the energy consumption decreases gradually for all the algorithms. Since increasing the number of learners can potentially allow orchestrators to associate with more learners, the learning task is distributed to more learners and each learner will have a smaller task size. Hence, the communication and computation energy consumption will be less for each learner. As for the accuracy, as shown in Fig. \ref{fig:my_eval_learners} (b), it starts to increase but then it decreases gradually as we increase the number of learners. In fact, smaller task sizes lead to less data transmission time between the orchestrator and the learners and less computation time at the learner, which can allow for more local iterations and global cycles. However, if the task size for each learner (i.e., the number of data samples) is small, it might not be sufficient to do the learning task and results in lower learning accuracy. On the other side, increasing the number of orchestrators means increasing the amount of data available for learning, which leads to a larger task size for each learner. As we can see in Fig. \ref{fig:my_eval_orchs} (a), the energy consumption increases at first due to the aforementioned fact, which leads to more computing and communication energy consumption. However, the task size for each learner can become large enough to throttle the learning process, since increasing the task size results in increasing the time needed for data transmission between the orchestrators and the learners, and more training time at the learner side. Therefore, the number of local iterations and global cycles will be decreased so that the total training time does not exceed its limit, which explains the sharp drop in the energy consumption after we increase the number of orchestrators. Similarly, in Fig. \ref{fig:my_eval_orchs} (b), we can see that the accuracy at first increases since larger task sizes for the learners result in a better learning experience. However, by limiting the number of local iterations and global cycles, the accuracy drops abruptly since the learners can not train sufficiently.

\subsection{Evaluation with Different Learning Tasks}
\begin{figure*}
    \centering
    \includegraphics[scale = 0.5]{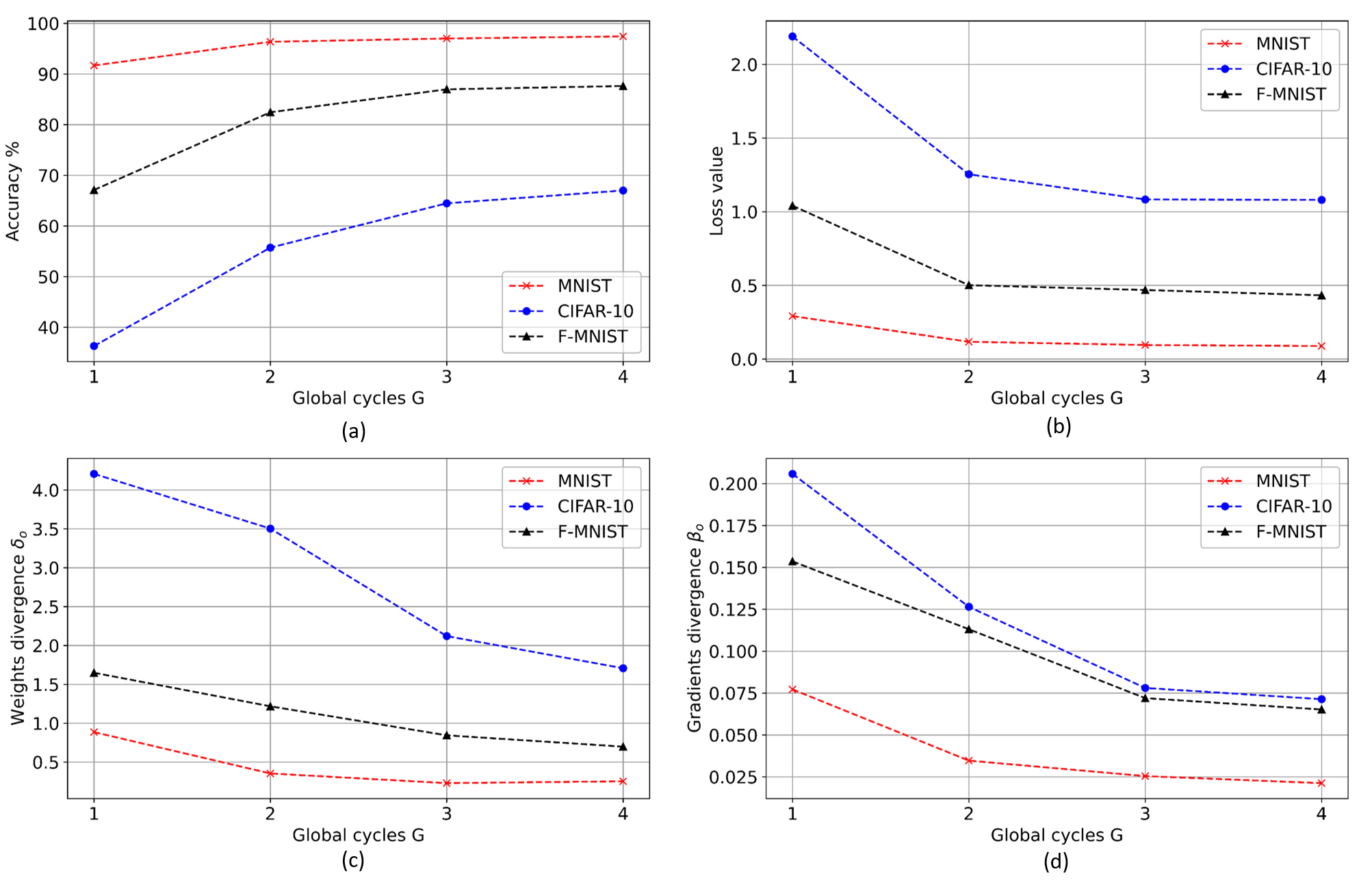}
    \caption{Learning metrics evaluation in terms of (a) global accuracy (b) global loss value (c) weights divergence (d) gradients divergence }
    \label{fig:learning_eval}
\end{figure*}

Lastly, we consider different multiple learning tasks with different datasets, namely, MNIST \cite{deng2012mnist}, FMNIST \cite{floudas2013deterministic}, and CIFAR-10 \cite{cifar}. We show the performance metrics for each learning task, specifically, the global accuracy, the global loss value, the weights, and the gradients divergence between the orchestrators' models and the associated learners' models. The global accuracy and loss value are plotted in Fig. \ref{fig:learning_eval} (a) and (b), respectively. We can notice for the MNIST and FMNIST datasets, the global models started to converge after 4 global cycles. Being a more complex learning task, the CIFAR-10 model did not converge within the same number of global cycles. However, if compared to the centralized training with the same number of total learning iterations which has an accuracy $\sim 73\%$, it only falls behind by $\sim 4\%$. As for the weights and gradients divergence, we claimed that these parameters can be empirically fixed to an upper bound to facilitate the analysis as provided in Table 1, where during training the actual values do not exceed this bound. The weights and gradients divergence are depicted in Fig. \ref{fig:learning_eval} (c) and (d), respectively. We noticed that during training, the values for both weights and gradients divergence are always below the upper bound for all the learning tasks, and the divergence gets smaller as the training progresses.

\subsection{Federated Learning Evaluation and Discussion}
\begin{figure}
    \centering
    \includegraphics[scale = 0.47]{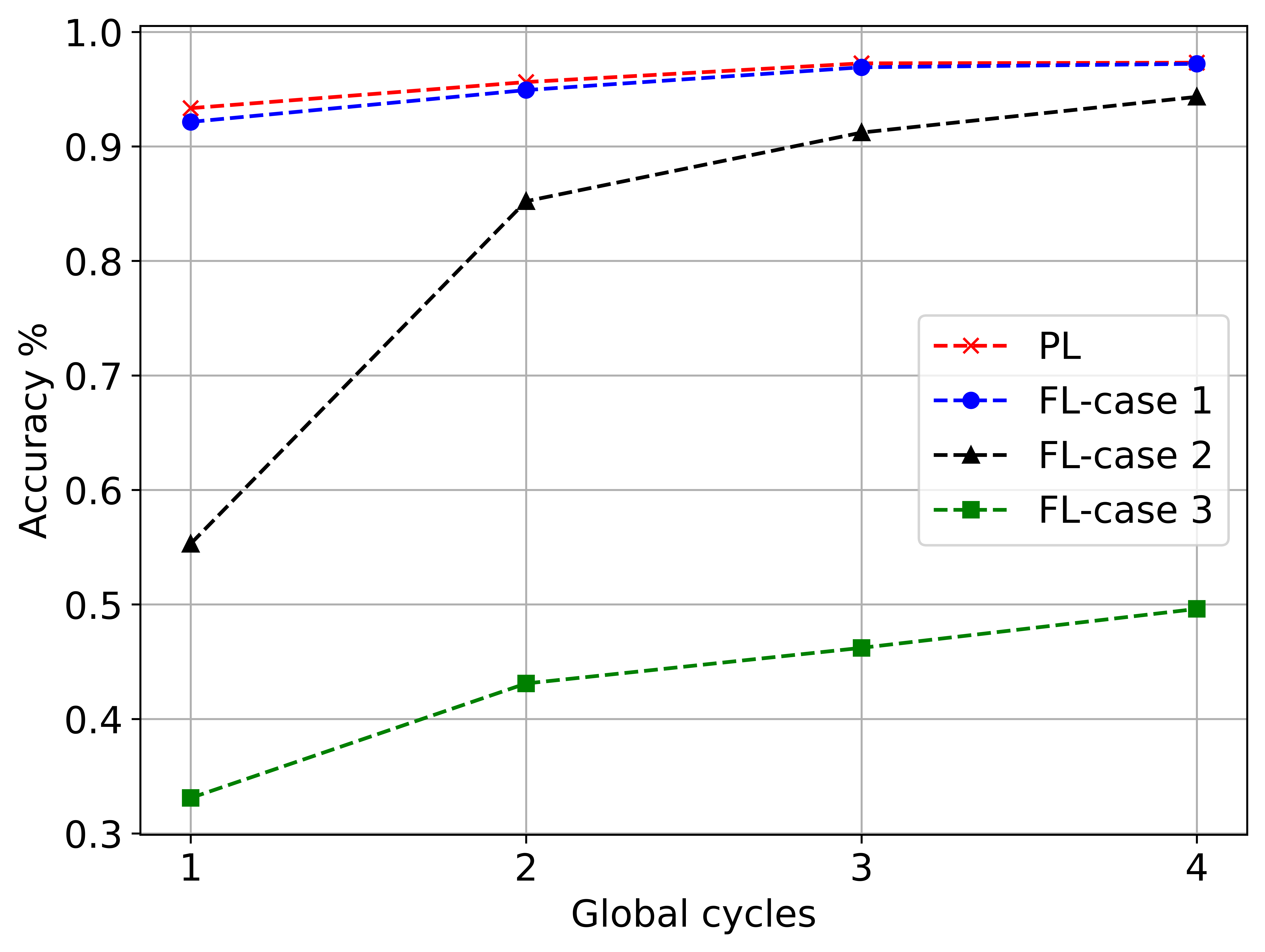}
    \caption{Accuracy evaluation of PL and FL with different cases.}
    \label{fig:PL_FL}
\end{figure}

In this section, we evaluate the FL in our model using the COPT approach for task allocation and FedAvg algorithm for distributed learning \cite{pmlr-v54-mcmahan17a} with different cases as follows: 1) \textbf{case 1}: The data is independent and identically distributed (IID) among all the learners, 2) \textbf{case 2}: The local datasets are not IID, with different amounts of data for each learner, 3) \textbf{case 3}: The data is completely non-IID and the distributions are skewed among learners. The accuracy evaluation of the FL cases compared to the PL is depicted in Fig \ref{fig:PL_FL}. It can be noticed that the FL with IID data performs similarly to the PL. In fact, since the orchestrator in the PL can control the data distribution, it can allocate the learning task to the learners such that the data is IID among them, which makes it identical to the FL in the first case. In the second case, it can be seen that the performance of FL drops at first, but as the training progresses the performance is improved and gets closer to the IID case. In the last case, it can be seen that the performance suffers from a sharp drop in accuracy with respect to the other cases. In fact, the FedAvg algorithm in the complete non-IID case fails to deliver any learning experience to the learners. Such downfall is already discussed in the literature \cite{Zhao2018,Wang2020}, and can be mitigated by improving the distributions of the learners' local datasets by sharing a small portion of data from each local dataset between the learners or selecting a subset of the associated learner to participate in the training. This can
come at the cost of violating the data privacy the main feature of FL, or missing learners with high capabilities that can speed up the training. 

\section{Conclusion}\label{sec:conc}
In this work, we studied the problem of an energy-aware, multi-task and multi-orchestrator MEL system. We first formulated a multi-objective optimization problem for learners-orchestrator association and task allocation that aims to minimize the total energy consumption and maximize the learning accuracy. Being NP-hard and non-convex problem, the problem is relaxed and covexified via exponential variable transformation and linear approximations for the non-convex terms, and then solved via the BnB algorithm.
Since the optimization approach is centralized and computationally expensive, we then proposed a set of lightweight partially and fully decentralized heuristics for the association and task allocation. The proposed heuristics let the orchestrators simply solve the association and the task allocation problems via convex optimization, and then determines the number of local iterations and global cycles via exhaustive search. To reduce the complexity and achieve a faster search, optimal upper bounds are derived for the number of local iterations and global cycles.
The conducted experiments show that the proposed approaches reduce the energy consumption significantly while executing multiple learning tasks compared to recent state-of-art methods while falling behind the benchmark by 2\%-3\% in terms of accuracy.

\section*{Acknowledgement}
This work was made possible by NPRP grant \# NPRP12S-0305-190231 from the Qatar National Research Fund (a member of Qatar Foundation). We also acknowledge the support of the Natural Sciences and Engineering Research Council of Canada (NSERC), [RGPIN-2020-06919]. The findings achieved herein are solely the responsibility of the authors.  

\printbibliography

\appendices

\section{}
\label{appA}
First, we define the separation function between the concave term and its under-estimator as follows:
\begin{equation}
\Delta(x) = -e^x - \left( - L(x)\,\right)
\end{equation}
\begin{equation}
 = -e^x + \left (  \frac{x_{max} e^{x_{min}} - x_{min} e^{x_{max}}} {x_{max} - x_{min}} + \frac{ e^{x_{max}} - e^{x_{min}}}{x_{max} - x_{min}} x   \right)
\end{equation}
The, we derive the first and second derivatives:
\begin{equation}
\frac{d \Delta }{dx} = -e^x + \frac{ e^{x_{max}} - e^{x_{min}}}{x_{max} - x_{min}}
\end{equation}
\begin{equation}
\frac{d^2 \Delta }{dx^2} = -e^x 
\end{equation}
It is readily obvious that $\Delta(x)$ is concave since its second derivative is always negative, and by setting $\frac{d \Delta }{dx}=0$, its maximum point can be given as:
\begin{equation}
x^* = \log\left(\frac{ e^{x_{max}} - e^{x_{min}}}{x_{max} - x_{min}} \right )
\end{equation}
by plugging it in the separation function, we can get the maximum separation value as follows:
\begin{equation}
\begin{split}
&\Delta(x^*) = -\frac{ e^{x_{max}} - e^{x_{min}}}{x_{max} - x_{min}} + \frac{x_{max} e^{x_{min}} - x_{min} e^{x_{max}}} {x_{max} - x_{min}}\\ 
&+ \frac{ e^{x_{max}} - e^{x_{min}}}{x_{max} - x_{min}} \times \log\left(\frac{ e^{x_{max}} - e^{x_{min}}}{x_{max} - x_{min}} \right ) 
\end{split}
\end{equation}
\begin{equation}
\begin{split}
&= e^{x_{min}} \bigg ( \frac{ 1 - e^{x_{max} - x_{min}}}{x_{max} - x_{min}} +  \frac{x_{max} - x_{min} e^{x_{max} - x_{min}} } {x_{max} - x_{min}} \\
& + \frac{ e^{x_{max} - x_{min}} -1 }{x_{max} - x_{min}} \log\left(\frac{  e^{x_{min}}  ( e^{x_{max} - x_{min} } - 1 )}{x_{max} - x_{min}} \right )  \bigg )
\end{split}
\end{equation}
By considering $\vartheta = x_{max} - x_{min}$ we can have:
\begin{equation}
\begin{split}
&= e^{x_{min}} \bigg ( \frac{ 1 - e^{\vartheta}}{\vartheta} +  \frac{x_{max} - x_{min} e^{\vartheta} + (e^{\vartheta} -1) x_{min}   } {\vartheta} \\
& + \frac{ e^{\vartheta} -1 }{\vartheta} \log\left(\frac{    e^{ \vartheta} - 1 }{\vartheta} \right )  \bigg )
\end{split}
\end{equation}
\begin{equation}
= e^{x_{min}} \bigg ( \frac{ 1 - e^{\vartheta}}{\vartheta} + 1 + \frac{ e^{\vartheta} -1 }{\vartheta} \log\left(\frac{    e^{ \vartheta} - 1 }{\vartheta} \right )  \bigg )
\end{equation}
Finally, by considering $Z = \frac{e^\vartheta - 1}{\vartheta}$, we can have:
\begin{equation}
\Delta(x^*) = \Delta_{max} = e^{x_{min}} \big( 1- Z + Z\log(Z) \big ) ~~~\qedsymbol
\end{equation}

\section{}
\label{appB}
The sub-problem \textbf{SP3} in (32) can be explicitly expressed as:
\begin{subequations}\small 
\begin{align}
&\underset{G_o, \tau_o}{\min.}\;\;\; \frac{\alpha}{E_{max} |\mathcal{L}_o| } \sum_{l \in \mathcal{L}_o} G_o (\zeta^2_{l,o} \tau_o n_{l,o} + \zeta^1_{l,o} n_{l,o} + \zeta^0_{l,o}  ) +  \frac{(1-\alpha)c1}{U_{max}\tau_o G_o} \\
&s.t. \;\;\;
G_o (A^2_{l,o} \tau_o n_{l,o} + A^1_{l,o} n_{l,o} + A^0_{l,o}  ) \leq T_{max}, \; \forall \;l \in \mathcal{L}_o\\
& \quad\;\; 1 \leq \tau_o \leq \tau_{max} \\
& \quad\;\; G_o \geq 1
\end{align}
\end{subequations}
Constraint (47b) represents the time constraints for each learner. However, we can substitute these constraint by a single one by considering the learner $l^*$ with the maximum training time  $l^* = \arg ~ \underset{l \in \mathcal{L}_o}{\max}~~ t_{l,o}$. Then, by using the following notations :
    \begin{description}
        \item $a = \frac{(1-\alpha)c1}{U_{max}},~~ b = \frac{\alpha  \sum_l \zeta^2_{l^*,o} n_{l^*,o} }{E_{max} |\mathcal{L}_o| }$
        \item $\\$
        \item $c = \frac{\alpha  \sum_l \left ( \zeta^1_{l^*,o} n_{l^*,o} + \zeta^1_{l^*,o} \right )}{E_{max} |\mathcal{L}_o| },~~ \theta= \frac{A^2_{l^*,o} n_{l^*,o}}{T_{max}}$
        \item $\\$
        \item $\xi =  \frac {(A^1_{l^*,o} n_{l^*,o} + A^0_{l^*,o}  )}{T_{max}}$
    \end{description}
The sub-problem \textbf{SP3} in (32) can be re-written as the following:
\begin{subequations}
\begin{align}
&\underset{G_o, \tau_o}{\min.}\;\;\; \frac{a}{\tau_o G_o} + b \tau_o G_o + c G_o \\
& s.t. \;\; \theta \tau_o G_o + \xi G_o \leq 1\\
& \quad\;\; 1 \leq \tau_o \leq \tau_{max} \\
& \quad\;\; G_o \geq 1
\end{align}
\end{subequations}
Afterwards, we assume the learner with the maximum training time takes his full time to train such that:
\begin{equation}
\theta \tau_o G_o + \xi G_o = 1
\end{equation}
So we can have the following equation:
\begin{equation}
\tau_o G_o =  \frac{1 - \xi G_o}{ \theta }
\end{equation}
By utilizing the above equation, the problem in (48) can be re-expresses as a single variable optimization problem $\mathcal{F}(G_o)$ as follows:
\begin{equation}
\min.\;\; \mathcal{F}(G_o) = \; \frac{a \theta}{1 - \xi G_o} + (c - \frac{b\xi}{\theta}) G_o
\end{equation}
\begin{equation}
s.t. \;\; G_o \geq 1
\end{equation}
Next, we derive the first and second derivatives of the objective function:
\begin{equation}
\frac{d \mathcal{F}}{G_o} = \frac{a\theta \xi}{(1-\xi G_o)^2} + c-\frac{b\xi}{\theta }
\end{equation}
\begin{equation}
\frac{d^2 \mathcal{F}}{G^2_o} = \frac{2a\theta \xi^2}{(1-\xi G_o)^3}
\end{equation}
From the above derivation, we can see that the second derivative is positive when $G_o < \frac{1}{\xi}$, and hence, the objective function is convex. The optimal point can be found by setting the first derivative to zero as follows:
\begin{equation}
\frac{a\theta \xi}{(1-\xi G^*_o)^2} + c-\frac{b\xi}{\theta } = 0
\end{equation}
\begin{equation}
\frac{a\theta^2 \xi}{(1-\xi G^*_o)^2} + c\theta -b\xi = 0
\end{equation}
\begin{equation}
(1-\xi G^*_o)^2 (c\theta -b\xi ) + a\theta^2 \xi = 0
\end{equation}
\begin{equation}
(1-\xi G^*_o)^2 = \frac{a\theta^2 \xi}{b\xi - c\theta }
\end{equation}
\begin{equation}
G^{*}_o = \left \lfloor \frac{1- \sqrt{\frac{\xi a \theta^2}{b\xi - \theta c}} }{\xi } \right \rfloor
\end{equation}
where we ignored the negative root since the value of $(1-\xi G^*_o)$ has to be positive, and floored the value since the number of global cycles is an integer. Moreover, one can see that for this solution to be feasible the following conditions must be satisfied  $\beta \xi - \theta c > \xi a \theta^2$. Lastly, the number of local training iterations can found using (50) and considering its maximum as follows:
\begin{equation}
\tau^{*}_o = \min \left ( \left \lfloor \frac{ 1 - \xi G^{*}_o  }{ \theta G^{*}_o } \right \rfloor, \tau_{max} \right ) ~~~\qedsymbol
\end{equation}

\section{}
\label{appC}
This appendix shows the Neural Networks architectures that have been used for each dataset. 
\subsection{MNIST and FMNIST}
Input(784) $\rightarrow$ Fully connected layer (784×256) $\rightarrow$ Activation layer (256) $\rightarrow$ Fully connected layer(256x256) $\rightarrow$ Activation layer (256) $\rightarrow$ Fully connected layer(256x10) $\rightarrow$ Softmax layer (10)
\subsection{CIFAR 10}
(Using the convention of ( (channels in, channels out), kernel size) for the convolutional layer), and convolutional layers contain internal activation layers)

Input(32x32x3) $\rightarrow$ Convolution (3,32), kernel (3,3) $\rightarrow$ Convolution (32,32), kernel (3,3) $\rightarrow$ Max pooling  layer (2,2) $\rightarrow$ Convolution (32,64), kernel (3,3) $\rightarrow$ Convolution (64,64), kernel (3,3) $\rightarrow$ Max pooling  layer (2,2) $\rightarrow$ Fully connected layer(256) $\rightarrow$ Activation layer (256) $\rightarrow$ Fully connected layer(256x10) $\rightarrow$ Softmax layer (10)
 
\end{document}